\documentclass[12pt]{article}
\usepackage{pdflscape}
\usepackage{amsmath,amssymb,amsthm,amsxtra,overpic,bbm,bm,epsfig,subfigure}
\usepackage{mathrsfs}
\usepackage{booktabs}
\usepackage{graphicx}
\usepackage{xcolor}
\usepackage{comment}
\usepackage{epstopdf}
\usepackage{float}
\usepackage{cite}
\usepackage{array}
\usepackage{makecell}
\usepackage{enumitem}
\usepackage{tabularx} 
\usepackage{overpic}
\usepackage{fancyhdr}
\usepackage{float}
\usepackage{rotating}
\usepackage{color}
\makeatletter
\usepackage{slashed,stmaryrd}
\def\thefootnote{\fnsymbol{footnote}}
\usepackage{geometry}
\definecolor{MyDarkBlue}{rgb}{0.1, 0.1, 0.8} 
\definecolor{SBlue}{rgb}{0.2, 0.4, 0.7} 
\definecolor{MyLightBlue}{rgb}{0.22,0.51,0.9}
\definecolor{MyGreen}{rgb}{0.0, 0.5, 0.0}
\definecolor{BrickRed}{rgb}{0.8, 0.25, 0.33}
\usepackage{hyperref}
\hypersetup{colorlinks, citecolor=SBlue,linkcolor=MyDarkBlue, urlcolor=BrickRed}

\begin{document}
\begin{center}
{\Large \bf 
Discovery potential of the Glashow resonance in an \\ air shower neutrino telescope
}
\end{center}
\renewcommand{\thefootnote}{\fnsymbol{footnote}}
\vspace{0.05in}
\begin{center}
{
	{}~\textbf{Guo-yuan Huang$^{1}$}\footnote{ E-mail: \textcolor{MyDarkBlue}{guoyuan.huang@mpi-hd.mpg.de}} 
}
\vspace{0.1cm}
{
	\\
	\em $^1$Max-Planck-Institut f{\"u}r Kernphysik, Saupfercheckweg 1, 69117 Heidelberg, Germany
} 
\end{center}
\renewcommand{\thefootnote}{\arabic{footnote}}
\setcounter{footnote}{0}
\thispagestyle{empty}
\vspace{0.5cm}
\begin{abstract}
\noindent The in-ice or in-water Cherenkov neutrino telescope such as IceCube has already proved its power in measuring the Glashow resonance by searching for the bump around $E^{}_{\rm \nu} = 6.3~{\rm PeV}$ arising from the $W$-boson production.
In the next few decades, there are many proposals that observe cosmic tau neutrinos with extensive air showers, also known as tau neutrino telescopes. As has been recognized, the air shower telescope is in principle sensitive to the Glashow resonance via the channel $W \to \tau \nu^{}_{\tau}$ followed by the tau decay in the air. However, with a thorough numerical analysis we have identified several limitations for those telescopes on hunting the resonance.
If ultrahigh-energy neutrinos are dominantly produced from the meson decay, 
it will be statistically difficult for a rather advanced proposal, such as TAMBO with a geometric area around $500~{\rm km^2}$, to discriminate the  Glashow resonance induced by $\overline{\nu}^{}_{e}$ from the intrinsic $\nu^{}_{\tau}/\overline{\nu}^{}_{\tau}$ background. 
The discovery significance is only around $1\sigma$ considering the flux parameters measured by IceCube as the input.
Nevertheless, the significance will be improved to $90\%$ if PeV neutrinos mainly originate from the neutron decay, which is, however, thought to be only a subdominant neutrino source. The presence of new physics can also increase the significance.
Compared to the in-ice or in-water telescope, the challenge for the Glashow resonance search is ascribed to several factors: (i) a suppressed branching ratio of $11\%$ for the decay $W \to \tau \nu^{}_{\tau}$; (ii) the smearing effect and the reduced acceptance because the daughter neutrino takes away $\langle y \rangle \sim 75\%$ of the energy from the $W$ decay; (iii) a large attenuation effect for Earth-skimming neutrinos with the resonance. 
\end{abstract}
\setcounter{footnote}{0}

\newpage

\section{Introduction}
In 1959, even before the theoretical establishment of the Standard Model (SM),  Glashow has proposed the resonant scattering of high-energy antineutrinos on the electron target via the weak interaction~\cite{Glashow:1960zz}, $\overline{\nu}^{}_{e} e \to W \to {\rm anything}$.
Almost sixty years later, the IceCube Observatory has finally identified such a candidate event with an energy deposition of $E^{}_{\rm dep} = 6.05 \pm 0.72~{\rm PeV}$~\cite{IceCube:2021rpz}, very close to the resonance energy at $E^{}_{\nu} \approx 6.3~{\rm PeV}$.
The fact that the cross section of resonance is higher than that of the deep inelastic scattering (DIS) by two orders of magnitude makes this event very likely 
to arise from the Glashow resonance process, i.e., with a statistical significance around $2\sigma$ confidence level (CL).

The observation of the Glashow resonance not only further strengthens the SM but also offers us a promising way to distinguish between different astrophysical neutrino sources~\cite{Berezinsky:1977sf,Brown:1981ns,Anchordoqui:2004eb,Hummer:2010ai,Xing:2011zm,Bhattacharya:2011qu,Bhattacharya:2012fh,Barger:2012mz,Barger:2014iua,Palladino:2015uoa,Shoemaker:2015qul,Anchordoqui:2016ewn,Kistler:2016ask,Biehl:2016psj,Sahu:2016qet,Huang:2019hgs,Zhou:2020oym,Bustamante:2020niz,Goncalves:2022uwy,Huang:2023yqz,Liu:2023lxz}.
Even though the IceCube Observatory has accumulated hundreds of neutrino events  with energies higher than $100~{\rm TeV}$ which should be of the astrophysical origin, the sources of those events are still largely unknown~\cite{IceCube:2013low,IceCube:2013cdw,IceCube:2018cha,IceCube:2020wum,IceCube:2022der,Abbasi:2023bvn}. 
Those ultrahigh-energy (UHE) neutrinos are thought to be produced by the scattering of accelerated cosmic rays with ambient photons or baryons surrounding the source, which can lead to characteristic $\overline{\nu}^{}_{e}$ fractions in the neutrino flux.
With the help of the Glashow resonance,
one can experimentally extract the $\overline{\nu}^{}_{e}$ fraction at Earth by comparing the resonance events to those induced by other neutrinos.
Recently, an analysis based on the candidate event of IceCube has been performed to infer the $\overline{\nu}^{}_{e}$ fraction at Earth~\cite{Huang:2023yqz}, and the sensitivities of next-generation in-ice and in-water Cherenkov telescopes have also been investigated~\cite{Liu:2023lxz}. 

Yet, another powerful category of UHE neutrino telescopes, which detects  extensive air showers induced by tau decays~\cite{Berezinsky:1975zz,Domokos:1997ve,Domokos:1998hz,Capelle:1998zz,Fargion:1999se,Fargion:2000iz,LetessierSelvon:2000kk,Feng:2001ue,Kusenko:2001gj,Bertou:2001vm,Cao:2004sd,Zas:2005zz,Baret:2011zz}, is firmly advancing (see Refs.~\cite{Huang:2021mki, Abraham:2022jse,Ackermann:2022rqc, Arguelles:2022xxa} for recent reviews).
The tau flux to detect is most efficiently produced through the charged-current (CC) conversion from tau neutrinos by scattering off  matter.
Thanks to the suitable decay length of tau above PeV energies, a large effective area can be achieved for neutrinos from PeV to ZeV depending on the specific layout of the telescope.
There have been a number of modern proposals towards this direction with an almost guaranteed discovery potential of cosmogenic tau neutrinos~\cite{Neronov:2016zou,GRAND:2018iaj,Otte:2018uxj,Otte:2019aaf,ARA:2019wcf,Abarr:2020bjd,Anker:2020lre,RNO-G:2020rmc,IceCube-Gen2:2020qha,POEMMA:2020ykm,Romero-Wolf:2020pzh,Wissel:2020fav,Wissel:2020sec,Hallmann:2021kqk,Ogawa:2021dK,deVries:2021BA}. 
In this regard, many theoretical studies have been performed to explore the particle physics potential of those facilities~\cite{Jezo:2014kla,Jho:2018dvt,Huang:2019hgs,Denton:2020jft,Soto:2021vdc,Huang:2021mki,Valera:2022ylt,Huang:2022pce,Esteban:2022uuw,Huang:2022ebg,Brdar:2022kpu,GarciaSoto:2022vlw,Valera:2022wmu,Heighton:2023qpg,Bertolez-Martinez:2023scp}.
Most of those telescopes optimize their sensitivities to the cosmogenic neutrino flux at EeV energies associated with the Greisen-Zatsepin-Kuzmin (GZK) cutoff structure of the cosmic ray spectrum~\cite{Greisen:1966jv,Zatsepin:1966jv,Beresinsky:1969qj}.
Their sensitivities usually decrease when going down to lower energies because of the detection threshold and the smaller acceptance with a shorter tau decay length.
%
%

Interestingly, an energy threshold down to $\mathcal{O}(\rm PeV)$ can be achieved by deploying the particle detector array along one side of a deep valley, monitoring the extensive air showers induced by tau decays from the other side, as in the Tau Air Shower Mountain-Based Observatory (TAMBO)~\cite{Romero-Wolf:2020pzh}.
TAMBO features the best sensitivity to tau neutrinos in the energy window from PeV to $100~{\rm PeV}$~\cite{Huang:2021mki}. 
A natural question arises: can this modern setup or other similar proposals (e.g., Ashra-NTA~\cite{Ogawa:2021dK}) with a low energy threshold identify the Glashow resonance at $E^{}_{\nu} \approx 6.3~{\rm PeV}$ induced by $\overline{\nu}^{}_{e}$? 

The channel $\overline{\nu}^{}_{e} e \to W \to \overline{\nu}^{}_{\tau} \tau$ followed by the tau decay in the air opens up the possibility.
Even though the branching ratio of $W \to \overline{\nu}^{}_{\tau} \tau$ is only $11\%$, the large cross section could in principle compensate the suppression.
Providing the excellent angular resolution, such facilities could create unique opportunities to extract the $\overline{\nu}^{}_{e}$ component even from the point source.
The idea of observing the Glashow resonance in the tau neutrino telescope was first noticed in the original paper by Fargion et al.~\cite{Fargion:1999se,Fargion:2000iz} and further addressed in Refs.~\cite{Zas:2005zz,GRAND:2018iaj,Huang:2019hgs,Soto:2021vdc,Liu:2023lxz} from various aspects~\footnote{In the air shower telescope, the Glashow resonance can be accessed via three possible ways~\cite{GRAND:2018iaj}: (i) the tau-decay air showers induced by $\overline{\nu}^{}_{e}$ interacting with mountain or Earth; (ii) the air showers induced by $\overline{\nu}^{}_{e}$ interacting with the atmosphere without the cosmic ray shielding; (iii) the shower leaked from the matter surface during development. 
However, the issue for the second one is that cosmic ray and gamma ray backgrounds are not shielded, while for the third one the shower intensity may be inefficient for detection due to the significant attenuation in rock.
Hence, the present work will focus on the typical tau-decay events with a clear detection efficiency.}. 
In light of the increasing interest of the community and fruitful measurements on the neutrino flux, we investigate the discovery potential of the Glashow resonance in the air shower neutrino telescope. 
By following a statistical framework,
we will provide  a rather quantitative analysis in this work.

The rest of the work is organized as follows. In Sec.~\ref{sec:II}, we describe the framework to calculate the event rates of the Glashow resonance induced by $\overline{\nu}^{}_{e}$ as well as the normal ${\nu}^{}_{\tau}/\overline{\nu}^{}_{\tau}$ events in the air shower telescope.
In Sec.~\ref{sec:III}, we present the event rates and distributions in a telescope similar to the TAMBO setup, and quantitatively compare the Glashow resonance signal to the ${\nu}^{}_{\tau}/\overline{\nu}^{}_{\tau}$ background.
We make our conclusion in Sec.~\ref{sec:IV}.

\section{Calculation framework} \label{sec:II}
The detection mechanism with air shower techniques relies on the suitable decay length of tau, $c \tau^{}_{\tau} \approx 50~{\rm km}\cdot ({E^{}_{\tau}/{\rm EeV}})$, making the tau flux able to emerge from the matter surface and decay to form extensive air showers.
The air shower can then be measured by detecting radio waves, Cherenkov light, fluorescence or directly shower particles, and those key technologies have been greatly developed and adopted in cosmic ray and gamma ray experiments.
One typical example is set by TAMBO~\cite{Romero-Wolf:2020pzh}, where the particle detection array will be deployed on one side of a deep valley (like a highly inclined AUGER), containing 22000 water tanks separated by $150~{\rm m}$ each.
By taking the valley length as $100~{\rm km}$ and the array width as $5~{\rm km}$, the total geometric area of the telescope is as large as $500~{\rm km}^2$. 
The Colca Valley of Peru with the sufficient depth and length is found to be an ideal site to host the proposed array.
The array will overwatch the other side of the valley by directly measuring shower particles induced by taus emerged from the matter surface.

The detection volume of the telescope depends not only on the geometric area of the telescope but also on the tau energy. 
Only those taus that can propagate out of the Earth before decay can be measured.
Generally speaking, the higher the tau energy is, the larger the effective volume can be for generating observable taus from the Earth. The  attenuation effect will also come into play when the cross section, which increases with the energy, is too large.
The actual detection volume would then be a joint result of the telescope geography, the tau energy as well as the matter attenuation effect.
For the deep-valley telescope, there are mainly four kinds of events depending on the incoming direction of tau, for which the primary neutrino has traveled through: (i) only mountain; (ii) only Earth; (iii) Earth and mountain; (iv) Earth, air and then mountain. Those that have traveled through the vast volume of Earth will be referred to as ``Earth-skimming neutrinos''. The attenuation effect for Earth-skimming neutrinos is much stronger than that for neutrinos only traveling through the mountain.

\begin{figure}
	\begin{center}
		\vspace{-0.6cm}
		\includegraphics[width=0.9\textwidth]{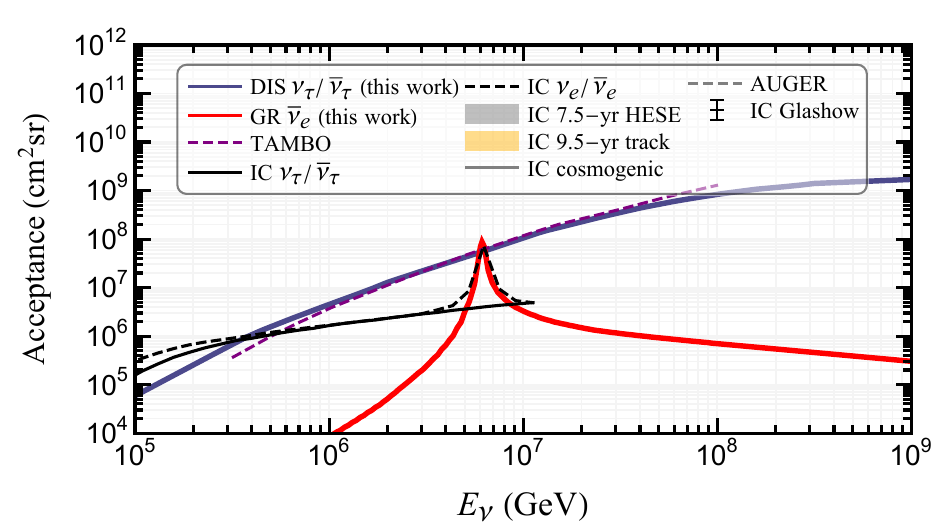}
		\includegraphics[width=0.9\textwidth]{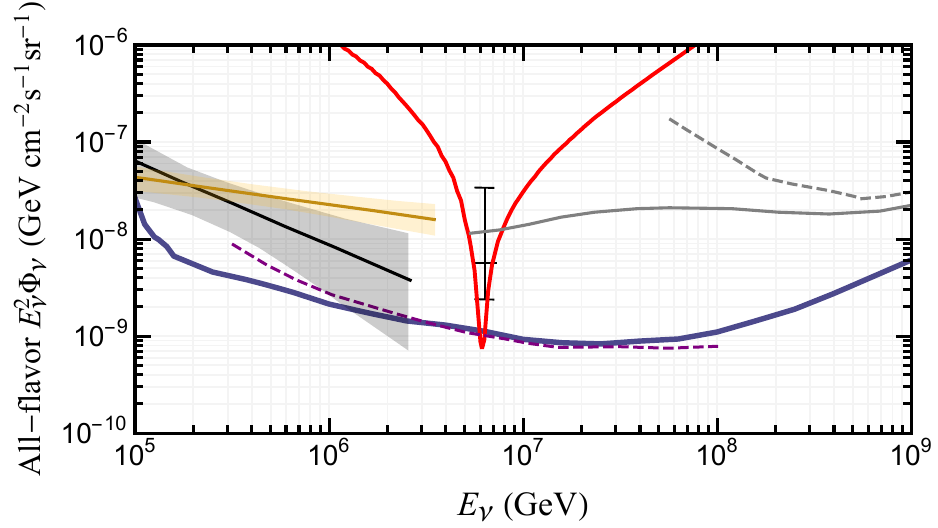}
	\end{center}
	\vspace{-0.3cm}
	\caption{The acceptance (upper panel) and 10-year sensitivity (lower panel) of our nominal neutrino telescope to $\nu^{}_{\tau}/\overline{\nu}^{}_{\tau}$ (dark blue curves) via DIS and to $\overline{\nu}^{}_{e}$ (red curves) via the Glashow resonance. This nominal setup follows closely the TAMBO proposal~\cite{Romero-Wolf:2020pzh}, whose original acceptance and sensitivity to $\nu^{}_{\tau}/\overline{\nu}^{}_{\tau}$ are given in purple curves. In the upper panel, acceptances of IceCube to $\nu^{}_{\tau}/\overline{\nu}^{}_{\tau}$ and $\overline{\nu}^{}_{e}$ are shown for comparison~\cite{IceCube:2013low}. In the lower panel, several measurements  on the diffuse neutrino neutrino flux are given, including: the IceCube 7.5-year HESE data~\cite{IceCube:2020wum}, 9.5-year northern sky track data~\cite{Stettner:2019tok}, one Glashow resonance candidate~\cite{IceCube:2021rpz}, and the constraints from the IceCube~\cite{IceCube:2018fhm} and AUGER~\cite{PierreAuger:2019ens} searches for cosmogenic neutrinos.}
	\label{fig:accep_sensi}
\end{figure}
%

For this study, we have developed our own numerical program to generate the events. It is worthwhile to mention other excellent codes available for public use such as ~\texttt{nuSQuIDS}~\cite{Arguelles:2020hss}, \texttt{NuTauSim}~\cite{Alvarez-Muniz:2018owm}, \texttt{NuPropEarth}~\cite{Garcia:2020jwr}, \texttt{nuPyProp}~\cite{NuSpaceSim:2021hgs}, \texttt{PROPOSAL}~\cite{Koehne:2013gpa} and \texttt{TauRunner}~\cite{Safa:2019ege,Safa:2021ghs} to solve the propagation of leptons in matter.
In our numerical calculation, we attempt to follow closely the configuration of TAMBO and simulate the neutrino propagation and the tau generation in matter. 
For the geography of the telescope, we adopt the cross sectional view in Fig.~5 of Ref.~\cite{Romero-Wolf:2020pzh} and assume a valley length of $100~{\rm km}$.
We further assume that the mountain is composed of the standard rock and the Earth follows the PREM profile~\cite{Dziewonski:1981xy}. 
The neutrino fluxes including $\nu^{}_{\tau}/ \overline{\nu}^{}_{\tau}$ and $ \overline{\nu}^{}_{e}$ are then injected isotropically.
The propagation equations in matter incorporating the Glashow resonance process read
\begingroup
\allowdisplaybreaks
\begin{align}
\frac{\mathrm{d}}{\mathrm{d} t} \left(\frac{\mathrm{d}\Phi^{}_{{\nu}^{}_{\tau} + \overline{\nu}^{}_{\tau}} }{ \mathrm{d} E^{}_{\nu}}  \right)  = & - N^{}_{\rm A} \rho \left(\sigma^{}_{\rm CC} + \sigma^{}_{\rm NC}  \right)  \frac{\mathrm{d}\Phi^{}_{{\nu}^{}_{\tau} + \overline{\nu}^{}_{\tau}} }{ \mathrm{d} E^{}_{\nu}}  +  N^{}_{\rm A} \rho \int \mathrm{d} E^{\prime}_{\nu}
\frac{\mathrm{d}\Phi^{}_{{\nu}^{}_{\tau} + \overline{\nu}^{}_{\tau}} }{ \mathrm{d} E^{\prime}_{\nu}}  \frac{1}{E^{\prime}_{\nu} } \left.\frac{\mathrm{d}\sigma^{}_{\rm NC}}{\mathrm{d} z}\right|_{z = \frac{E^{}_{\nu}}{E^{\prime}_{\nu}} } \\
& + N^{}_{\rm A} \rho \frac{Z}{A} {\rm Br}^{}_{W\to \tau \overline{\nu}^{}_{\tau}} \int \mathrm{d} E^{\prime}_{\nu}
\frac{\mathrm{d}\Phi^{}_{\overline{\nu}^{}_{e}} }{ \mathrm{d} E^{\prime}_{\nu}}  \frac{1}{E^{\prime}_{\nu} } \left.\frac{ \mathrm{d}\sigma^{}_{\rm GR}}{\mathrm{d} z}\right|_{z = \frac{E^{}_{{\nu}}}{E^{\prime}_{\nu}} } 
+ 
\int \mathrm{d} E^{\prime}_{ \tau} \frac{\mathrm{d} \Phi^{}_{\tau}}{\mathrm{d} E^{\prime}_{ \tau}} 
\frac{1}{E^{\prime}_{\tau}} \frac{\mathrm{d} \Gamma^{}_{\tau} }{ \Gamma^{}_{\tau} \mathrm{d} z} \;, \notag
\\ 
\frac{\mathrm{d}}{\mathrm{d} t} \left(\frac{\mathrm{d}\Phi^{}_{\overline{\nu}^{}_{e}} }{ \mathrm{d} E^{}_{\nu}}  \right)  = & - N^{}_{\rm A} \rho \left(\sigma^{}_{\rm CC} + \sigma^{}_{\rm NC} + \frac{Z}{A} \sigma^{}_{\rm GR} \right)  \frac{\mathrm{d}\Phi^{}_{\overline{\nu}^{}_{e}} }{ \mathrm{d} E^{}_{\nu}}  +  N^{}_{\rm A} \rho \int \mathrm{d} E^{\prime}_{\nu}
\frac{\mathrm{d}\Phi^{}_{\overline{\nu}^{}_{e}} }{ \mathrm{d} E^{\prime}_{\nu}}  \frac{1}{E^{\prime}_{\nu} } \left.\frac{\mathrm{d}\sigma^{}_{\rm NC}}{\mathrm{d} z}\right|_{z = \frac{E^{}_{\nu}}{E^{\prime}_{\nu}} } \\
& + N^{}_{\rm A} \rho \frac{Z}{A} {\rm Br}^{}_{W\to e \overline{\nu}^{}_{e}} \int \mathrm{d} E^{\prime}_{\nu}
\frac{\mathrm{d}\Phi^{}_{\overline{\nu}^{}_{e}} }{ \mathrm{d} E^{\prime}_{\nu}}  \frac{1}{E^{\prime}_{\nu} } \left.\frac{ \mathrm{d}\sigma^{}_{\rm GR}}{\mathrm{d} z}\right|_{z = \frac{E^{}_{\nu}}{E^{\prime}_{\nu}} }  \;,\notag
\\ 
\frac{\mathrm{d}}{\mathrm{d} t} \left(\frac{\mathrm{d}\Phi^{}_{\tau} }{ \mathrm{d} E^{}_{\tau}}  \right)  = & - \Gamma^{}_{\tau} \frac{\mathrm{d}\Phi^{}_{\tau} }{ \mathrm{d} E^{}_{\tau}} 
-  \frac{N^{}_{\rm A}\rho}{A} \sigma^{}_{\rm \tau} \frac{\mathrm{d}\Phi^{}_{\tau} }{ \mathrm{d} E^{}_{\tau}}   
+
\frac{N^{}_{\rm A}\rho}{A}\int \mathrm{d} E^{\prime}_{\tau} \frac{\mathrm{d}\Phi^{}_{\tau}}{\mathrm{d} E^{\prime}_{\tau}} \frac{1}{E^{\prime}_{\tau}} 
\frac{\mathrm{d} \sigma^{}_{\rm \tau} }{\mathrm{d} z} 
+
\rho \frac{\partial}{\partial E^{}_{\tau}} \left(  \beta^{}_{\tau}  E^{}_{\tau} \frac{\mathrm{d}{\Phi^{}_{\tau}}}{\mathrm{d} E^{}_{\tau}}   \right) 
\\
&  +  N^{}_{\rm A} \rho \int \mathrm{d} E^{\prime}_{\nu} \frac{\mathrm{d} \Phi^{}_{{\nu}^{}_{\tau} + \overline{\nu}^{}_{\tau}}}{\mathrm{d} E^{\prime}_{\nu}} \frac{1}{ E^{\prime}_{ \nu }} \frac{\mathrm{d}\sigma^{}_{\rm CC}}{\mathrm{d} z}  
+ 
N^{}_{\rm A} \rho \frac{Z}{A} {\rm Br}^{}_{ W\to \tau \overline{\nu}^{}_{\tau} } \int \mathrm{d} E^{\prime}_{\nu}
\frac{\mathrm{d}\Phi^{}_{\overline{\nu}^{}_{e}} }{ \mathrm{d} E^{\prime}_{\nu}}  \frac{1}{E^{\prime}_{\nu} } \left.\frac{ \mathrm{d}\sigma^{}_{\rm GR}}{\mathrm{d} y}\right|_{y = \frac{E^{}_{\tau}}{E^{\prime}_{\nu}} }\;, \notag
\end{align}
\endgroup
where $N^{}_{\rm A}$ is the Avogadro constant, $\rho$ is the matter density, $Z$ is the atomic number, $A$ is the mass number, $\sigma^{}_{\rm CC}$ and $\sigma^{}_{\rm NC}$ represent the DIS cross sections of the charged-current and neutral-current (NC) interactions, respectively, and $\Gamma^{}_{\tau}$, $\sigma^{}_{\tau}$ and $\beta^{}_{\tau}$ describe the decay and the attenuation of tau in matter.
Here, $\sigma^{}_{\rm GR}$ stands for the cross section of the Glashow resonance, which takes a Breit-Wigner form. ${\rm Br}^{}_{W\to\tau \overline{\nu}^{}_{\tau}} \approx {\rm Br}^{}_{W\to e \overline{\nu}^{}_{e}} \approx 11\%$ are the branching ratios of the decay channels of interest.
The differential cross section of the resonant scattering reads $\mathrm{d}\sigma^{}_{\rm GR}/{\mathrm{d} y} = \sigma^{}_{\rm GR} \cdot 3  (1-y)^2 $ with $y \equiv E^{}_{\tau} / E^{\prime}_{\nu}$ being the energy fraction of the daughter tau taken from the primary neutrino~\cite{Zas:2005zz}.
Note that the secondary $\nu^{}_{e}$ and $\nu^{}_{\mu}$ from tau decays are not included in the above equations, which only have a subleading effect~\cite{Beacom:2001xn,Dutta:2002zc,Safa:2019ege,Safa:2021ghs}.

Due to the angular momentum conservation, the average fraction of energy carried by the daughter tau from $W$ is suppressed, i.e., only around $\langle y \rangle \sim 25\%$ in comparison with $\langle y \rangle \sim 80\%$ for the DIS CC interaction. This will reduce the detection volume of the Glashow resonance by a factor of three, as compared to the normal tau neutrino events for the same primary neutrino energy.
The detection volume will be further reduced due to the large cross section, e.g., $\sigma^{\rm max}_{\rm GR} \approx 5 \times 10^{-31}~{\rm cm^2}$, corresponding to an attenuation length of $L^{}_{\rm GR} \approx 30 ~{\rm km}$ in the standard rock.
Hence, the attenuation effect will greatly decrease the event number for the upward going Earth-skimming neutrinos, which traverse a distance much longer than $L^{}_{\rm GR}$.

By solving the propagation equations, one can obtain the tau flux emerged from the matter surface and then simulate their decays in the air.
For simplicity, we assume that the detector will be triggered as long as the air shower axis intersects with the area covered by the array.
Note that this is a rather optimistic assumption especially for less energetic neutrinos. In practice, one has to perform the Monte-Carlo simulation for the development of extensive air showers and the event registration in water tanks. The detector may not be triggered if the shower energy is too low.
Nonetheless, our result presented here should be regarded as the most optimistic limit which might be reached by further improving the detection threshold. We will show later that it is still statistically not easy to identify the Glashow resonance in such an ideal case. Incorporating the practical shower simulation should further reduce the discovery potential.

Using the framework above, we can first check the sensitivity of our nominal telescope to the diffuse neutrino flux.
In Fig.~\ref{fig:accep_sensi}, we show the acceptance (upper panel) and the sensitivity (lower panel) at $90\%$ CL for $\nu^{}_{\tau}/\overline{\nu}^{}_{\tau}$ (dark blue curves) via DIS and $\overline{\nu}^{}_{e}$ (red curves) via the Glashow resonance based on our simulation. 
The sensitivity to the all-flavor neutrino flux is obtained by requiring the differential event number per energy decade is greater than 2.44 for $90\%$ CL and assuming a democratic flavor composition along with $\nu^{}_{e}:\overline{\nu}^{}_{e} = 0:1$.
For comparison, the dashed purple curve stands for the acceptance and the corresponding sensitivity of the original TAMBO proposal to $\nu^{}_{\tau}/\overline{\nu}^{}_{\tau}$~\cite{Romero-Wolf:2020pzh}. 
One can notice that the result of our ideal simulation is quite close to the TAMBO curve and starts to deviate considerably when the neutrino energy is below PeV.
This deviation indicates that the energy threshold of TAMBO becomes  important  for $E^{}_{\nu} \lesssim \mathcal{O}(\rm PeV)$, above which the detector can accept tau events with a nearly full efficiency.
The sensitivity to $\overline{\nu}^{}_{e}$ with the Glashow resonance can exceed that to $\nu^{}_{\tau}/\overline{\nu}^{}_{\tau}$ in a narrow energy window around $E^{}_{\nu} \approx 6.3~{\rm PeV}$. However, the actual event number from the Glashow resonance is subject to an integration over the input neutrino flux.

For comparison, several existing results have been shown: the IceCube sensitivity to $\nu^{}_{\tau}/\overline{\nu}^{}_{\tau}$ (solid black curve) and $\overline{\nu}^{}_{e}$ (dashed black curve) via both DIS and the Glashow resonance~\cite{IceCube:2013low}, the IceCube measurements on the diffuse neutrino flux with 7.5-year HESE data (gray band)~\cite{IceCube:2020wum} and 9.5-year through-going track data (orange band)~\cite{Stettner:2019tok}, the flux estimate from the IceCube Glashow resonance candidate (black error bar)~\cite{IceCube:2021rpz}, and the cosmogenic neutrino searches by IceCube (solid gray curve)~\cite{IceCube:2018fhm} and AUGER (dashed gray curve)~\cite{PierreAuger:2019ens}. 
Notice that TAMBO's sensitivity  to $\nu^{}_{\tau}/\overline{\nu}^{}_{\tau}$ is better than that of IceCube by more than one order of magnitude above $\mathcal{O}(10~{\rm PeV})$. However, its sensitivity to $\overline{\nu}^{}_{e}$ via the Glashow resonance is almost comparable to that of IceCube, due to the aforementioned suppression of the detection volume.


\section{The Glashow resonance events}  \label{sec:III}
We continue with investigating the potential of air shower neutrino telescopes to the Glashow resonance by comparing event numbers and distributions of both the signal and the background.
As has been mentioned, the major background comes from the irreducible tau neutrino events, which have the same event topology as the Glashow resonance signal in the air shower neutrino telescope.
The only feasible way to single out the Glashow resonance seems to be statistically analyzing event distributions.
Owing to neutrino oscillations, there will always be a comparable tau neutrino component no matter what the initial neutrino flavor ratio is at source~\cite{Mena:2014sja, Chen:2014gxa, Palomares-Ruiz:2015mka, Aartsen:2015ivb, Palladino:2015zua, Arguelles:2015dca, Bustamante:2015waa, Aartsen:2015knd, Brdar:2016thq, DAmico:2017dwq, Pagliaroli:2015rca, Rasmussen:2017ert, Brdar:2018tce, Bustamante:2019sdb, Palladino:2019pid, Stachurska:2019srh,Song:2020nfh}.
For instance, the initial flavor ratios for the proton-proton (pp) and proton-photon (p$\gamma$) sources are both around $\nu^{}_{e} + \overline{\nu}^{}_{e} : \nu^{}_{\mu} + \overline{\nu}^{}_{\mu} : \nu^{}_{\tau} + \overline{\nu}^{}_{\tau} = 1:2:0$, which leads to a nearly democratic flavor composition $\approx 1:1:1$ at Earth.
In following demonstrations, without otherwise specified the flavor ratio will be set to $ 1:1:1$ with an optimistic input $\nu^{}_{e}:\overline{\nu}^{}_{e} = 0:1$.

\begin{figure}[t!]
	\begin{center}
		\vspace{-0.3cm}
		\includegraphics[width=0.45\textwidth]{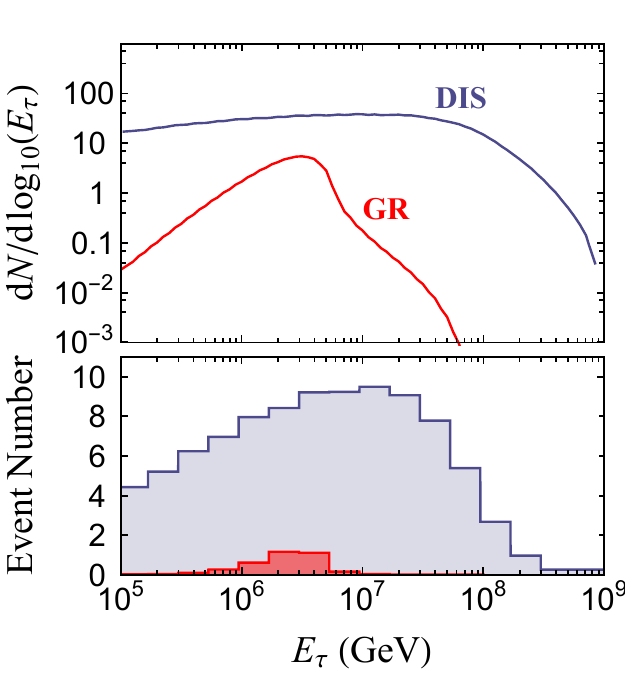}
		\hspace{0.3cm}
		\includegraphics[width=0.45\textwidth]{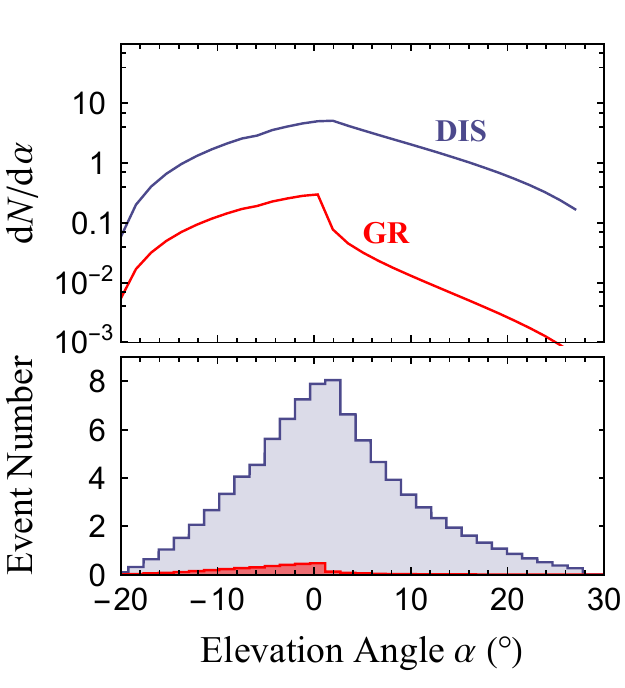}
	\end{center}
	\vspace{-1cm}
	\caption{The event distributions (upper panels) and numbers (lower panels) in terms of the tau energy $E^{}_{\tau}$ (left panel) and the incoming elevation angle $\alpha$ (right panel), assuming ten years of data collection. The Glashow resonance and tau neutrino events are given in red and blue, respectively. A power-law neutrino spectrum has been taken as Eq.~(\ref{eq:dPhidE}) with ${\rm \Phi}= 4.32$ and $\gamma = 2.28$.}
	\label{fig:eventDist}
\end{figure}
In the in-ice or in-water Cherenkov telescope such as IceCube, the Glashow resonance stands out as a hadronic cascade with an energy deposition close to $E^{}_{\rm dep} \sim 6.3~{\rm PeV}$. In the narrow energy window of interest, the background contribution from DIS is very low.
However, this is not the case for the air shower neutrino telescope, where most of the incoming neutrino energy does not result into the cascade.
The resonance bump is largely smoothed out due to the wide spread of the tau energy from $0~{\rm PeV}$ to $6.3~{\rm PeV}$.
To see it explicitly, in Fig.~\ref{fig:eventDist} we show the energy (left panel) and angular (right panel) distributions of tau events in the telescope with ten years of exposure.
The upper panels stand for the differential event distributions in the logarithmic scale and the lower ones give the event numbers collected in each bin.
The red curves and regions depict the Glashow resonance events while the blue ones are for the tau neutrino events.
For illustration, we have taken a power-law neutrino spectrum as the input:
\begin{align} \label{eq:dPhidE}
\frac{\mathrm{d} {\rm \Phi}^{}_{6\nu}}{\mathrm{d} E^{}_{\nu}} = {\rm \Phi} \left(\frac{E^{}_{\nu}}{100~{\rm TeV}}\right)^{-\gamma}10^{-18}~ {\rm GeV^{-1} cm^{-2} s^{-1} sr^{-1}} \;,
\end{align}
with ${\rm \Phi}= 4.32$ and $\gamma = 2.28$, i.e., the best-fit values of IceCube's 9.5-year track data~\cite{Stettner:2019tok}.
The elevation angle in the right panel is defined as $\alpha \equiv \theta-90^{\circ}$, where $\theta$ is the zenith angle of neutrino's incoming direction with $z$-axis  pointing upwards.
Further remarks on Fig.~\ref{fig:eventDist} are made below.
\begin{itemize}[noitemsep,topsep=0pt,leftmargin=5.5mm]
	\item For the chosen flux parameters, the expectation number for the Glashow resonance events with ten years of data collection is found to be $N^{}_{\rm GR} \approx 2.9$, which is promising to observe if we could neglect the background (i.e., tau neutrinos).
	The background event number shall be counted within a given region of interest (ROI).
	As shown in the left panel of Fig.~\ref{fig:eventDist}, even though the resonance cross section features a very narrow width, the energy distribution of tau events spreads over a wide range. 
	We find that the event number of the Glashow resonance is peaked around $E^{}_{\tau} \approx 2.2~{\rm PeV}$ with a large standard deviation in the energy $\Delta \ln(E^{}_{\tau}) \equiv \Delta E^{}_{\tau}/E^{}_{\tau} \approx  70\%$, corresponding to an energy interval of $E^{}_{\tau} \in \left[1.1,\; 4.5\right]~{\rm PeV}$.
	This energy interval is suitable to be considered as the ROI in order to optimize the signal-to-background ratio: for a wider interval one may involve too many background events compared to the signal, while for a narrower interval the signal events may not be sufficiently taken into account.
	\item On the other hand, the angular distribution in the right panel motivates us to impose a cut $\alpha< 2^{\circ}$ to reduce the background. 
	As we have mentioned, the Glashow resonance events for Earth-skimming neutrinos ($\alpha \gtrsim 2^{\circ}$) are strongly suppressed due to attenuation. 
	The angular cut will reduce the background event number by approximately a fact of two, while the signal is almost unchanged.
	After the angular cut, the background event number within the ROI turns out to be $N^{}_{\rm DIS} \approx 10$ with a $1\sigma$ fluctuation $\sqrt{N^{}_{\rm DIS}} \approx 3.2$, which is just comparable to $N^{}_{\rm GR}$. This implies that the signal significance is only around $1\sigma$ CL for the assumed flux input.
	\item
	In practice, the ROI choice is further limited by the energy resolution of taus in the telescope.
	For the TAMBO proposal, the target energy resolution is around $\Delta E^{}_{\tau}/E^{}_{\tau} \approx  100\%$~\cite{Romero-Wolf:2020pzh} which will contribute an additional smearing effect on top of the intrinsic energy spread.
\end{itemize}

\begin{figure}
	\begin{center}
		\vspace{-1cm}
		\includegraphics[width=0.45\textwidth]{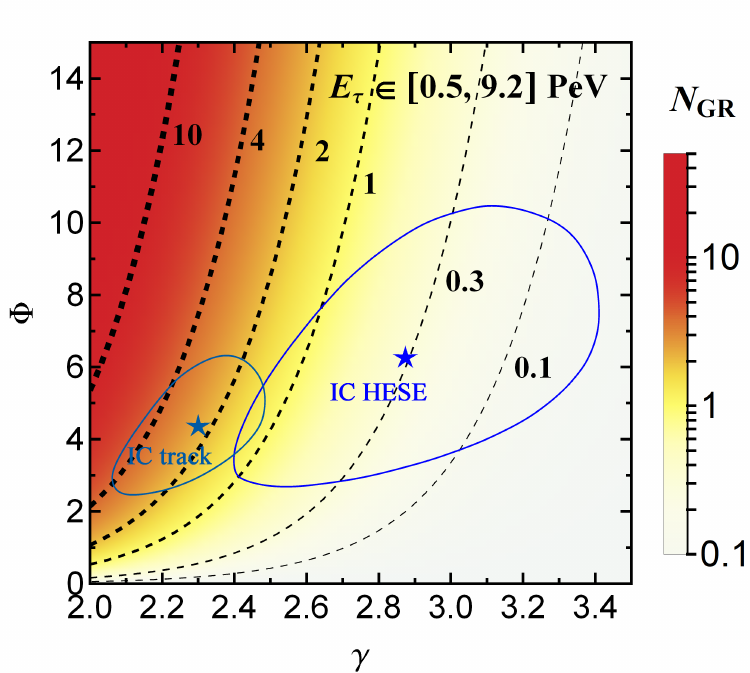}
		\includegraphics[width=0.45\textwidth]{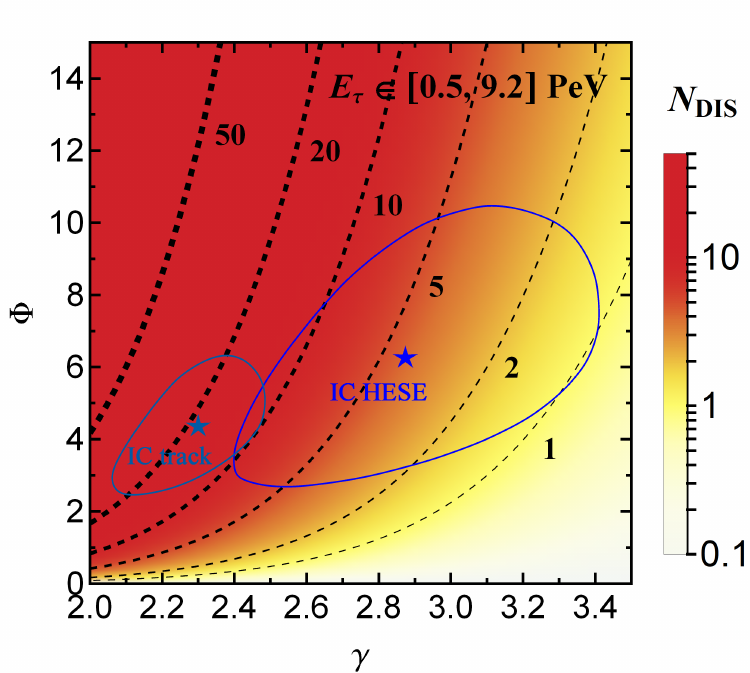}
		\includegraphics[width=0.45\textwidth]{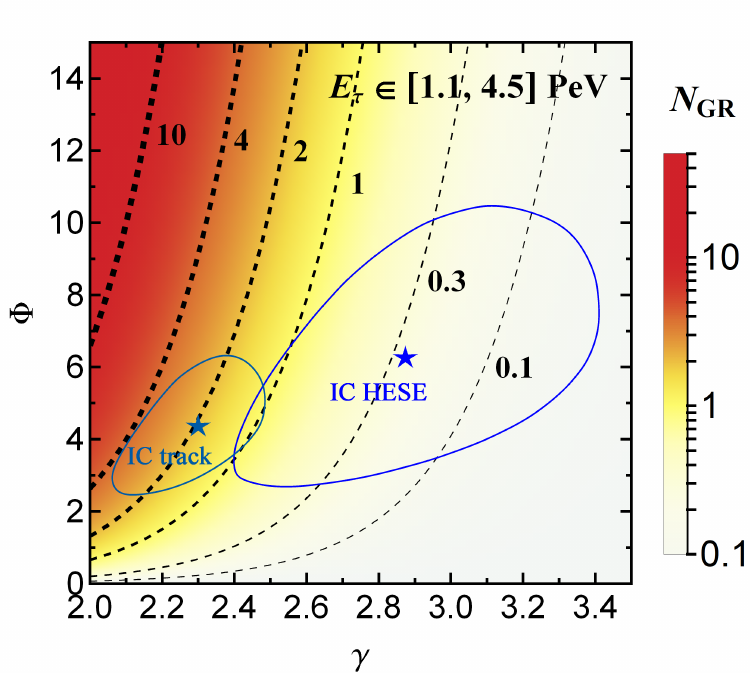}
		\includegraphics[width=0.45\textwidth]{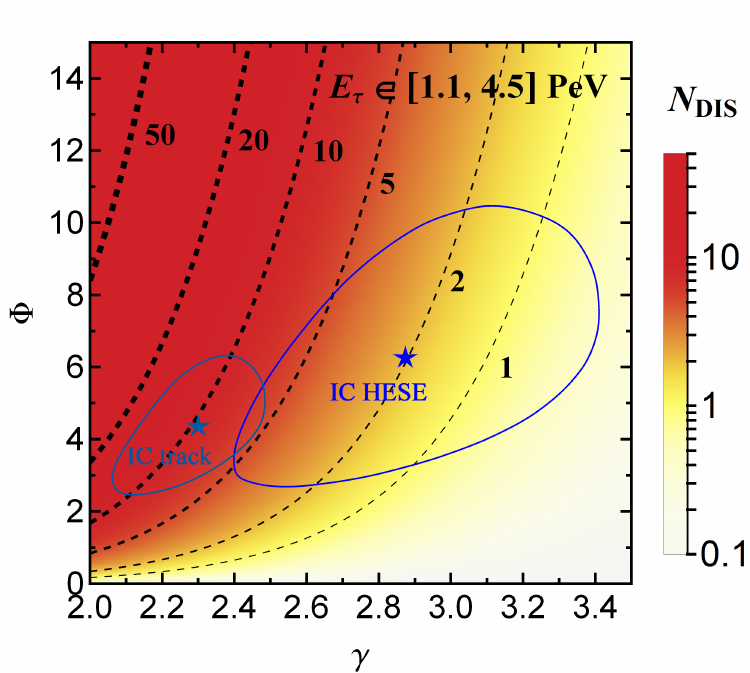}
		\includegraphics[width=0.45\textwidth]{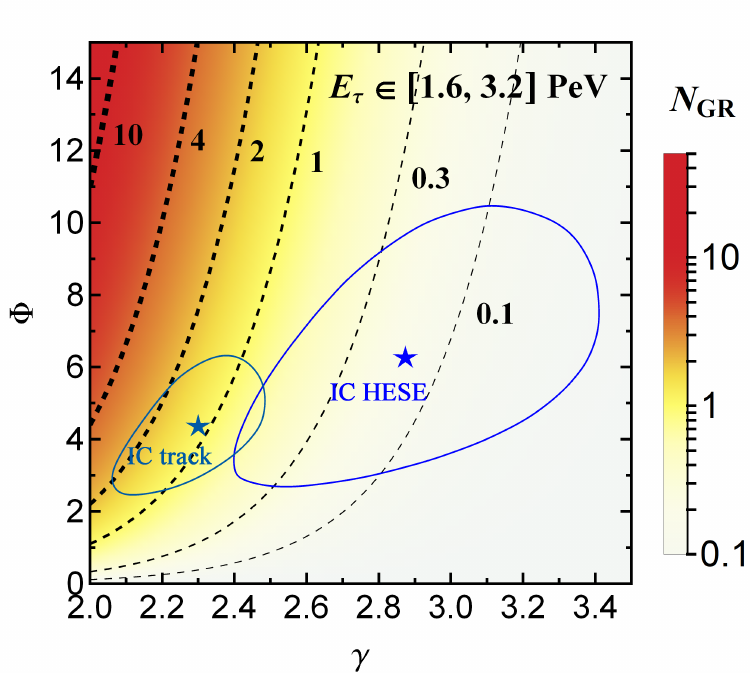}
		\includegraphics[width=0.45\textwidth]{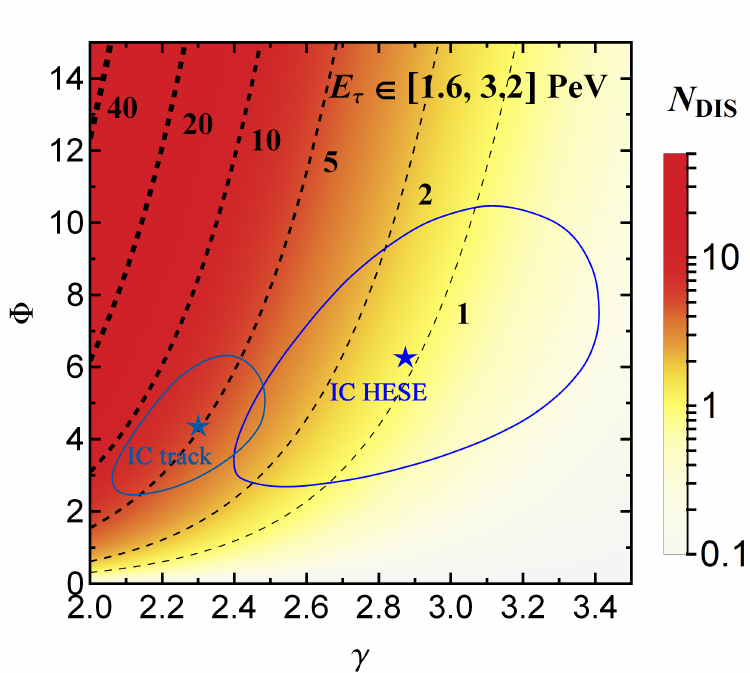}
	\end{center}
	\vspace{-0.3cm}
	\caption{The event numbers within different energy intervals through the Glashow resonance (left panels) and DIS (right panels), assuming ten years of data collection. The parameter regions preferred by IceCube's HESE and track data are shown as the solid contours (at $2\sigma$ CL) and stars (best fits) for comparison.}
	\label{fig:eventContour}
\end{figure}

For the above example, we have fixed the flux normalization $\Phi$ and spectral index $\gamma$ to be the best-fit values of IceCube track data. 
In order to be more general, we now set them as free parameters and perform the scanning.
The event numbers of the signal and background in terms of $\Phi$ and $\gamma$ are given in Fig.~\ref{fig:eventContour}.
In each panel, the parameter regions preferred by IceCube's HESE~\cite{IceCube:2020wum} and through-going track data~\cite{Stettner:2019tok} are shown as the solid contours (at $2\sigma$ CL) and stars (best fits).
To maximize the signal-to-background ratio, the angular cut $\alpha <2^{\circ}$ has been imposed.
For illustration, from top to bottom panels we have chosen three different ROIs while calculating the event numbers, including $E^{}_{\tau} \in \left[0.5,\; 9.2 \right]~{\rm PeV}$, $\left[1.1,\; 4.5 \right]~{\rm PeV}$ and $\left[1.6,\; 3.2 \right]~{\rm PeV}$, which correspond to multiplying the standard deviation of the tau energy distribution by a factor of two, one and half, respectively. 
For a large fraction of the parameter space favored by HESE data, the event number of the Glashow resonance is small, $N^{}_{\rm GR} \lesssim 1$.
For the parameter space favored by track data, a larger event number can be obtained, but the ratio $N^{}_{\rm GR}/\sqrt{N^{}_{\rm DIS}} \lesssim 1$ is not sufficient to claim that the signal is significant against the background.

When the flux normalization is too small or the spectrum is too soft, $N^{}_{\rm DIS}$  within the ROI can be of $\mathcal{O}(1)$, in which case $N^{}_{\rm GR}/\sqrt{N^{}_{\rm DIS}}$ is no longer a good estimate for the signal significance. 
In order to accommodate all cases, we shall treat the event fluctuation according to the Poisson distribution, for which the probability distribution function (PDF) follows ${\rm PDF}^{}_{\rm Poisson} (n, \mu) = e^{-\mu} \mu^n / {n!}$, where $\mu$ is the expectation value and $n$ is the number of counts. The corresponding cumulative distribution function (CDF) is defined as ${\rm CDF}^{}_{\rm Poisson} (n, \mu)  \equiv \sum^n_{k = 0} {\rm PDF}^{}_{\rm Poisson} (k, \mu)$, and $\overline{\rm CDF}^{}_{\rm Poisson} = 1 - {\rm CDF}^{}_{\rm Poisson}$ is its complementary function.
To define the discovery probability of the signal, we can perform a set of pseudo-experiments, and calculate the $p$-value that a fraction of pseudo-experiments can exclude the null (background-only) hypothesis.  
The $p$-value can be figured out by solving the following relations~\cite{Punzi:2003bu}:
\begin{eqnarray}\label{eq:ses}
p = {\rm CDF}^{}_{\rm Poisson} (n^{}_{p }, N^{}_{\rm DIS})  \; ,~
\overline{\rm CDF}^{}_{\rm Poisson} (n^{}_{p }, N^{}_{\rm DIS} + N^{}_{\rm GR})  = q   \; ,
\end{eqnarray}
where $n^{}_{p }$ is the mock event number that can be observed by a fraction $q$ of pseudo-experiments including both signal and background events. 
The median sensitivity corresponding to $q = 50\%$ will be taken, so that $p$ will quantify the significance that half of pseudo-experiments can exclude the null hypothesis.
For illustration, we have smoothed over the CDF by taking ${\rm CDF}^{}_{\rm Poisson} (x, \mu) = \Gamma(x+1,\mu)/\Gamma(x+1)$ with $x \in \mathbb{R}$ following Ref.~\cite{Agostini:2017jim}.

\begin{figure}[t!]
	\begin{center}
		\vspace{-0.3cm}
		{	\includegraphics[width=0.28\textwidth]{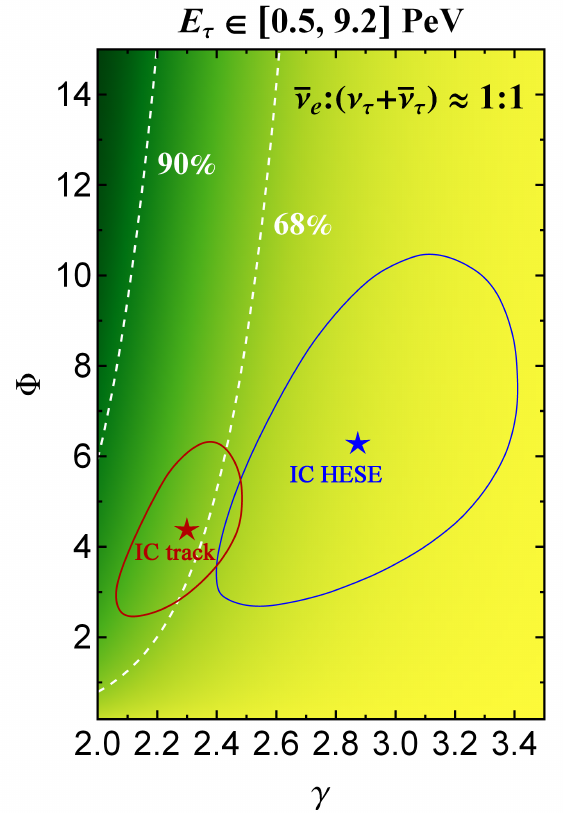}
			\includegraphics[width=0.28\textwidth]{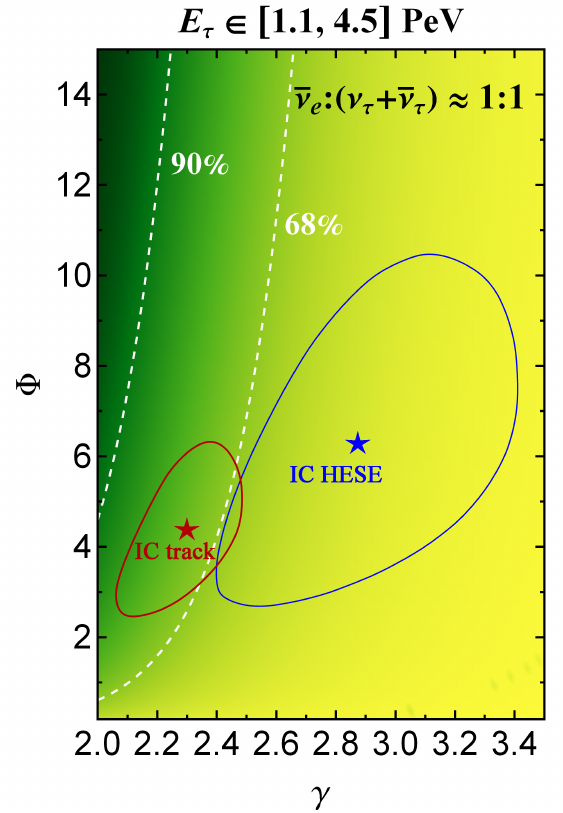}
			\includegraphics[width=0.344\textwidth]{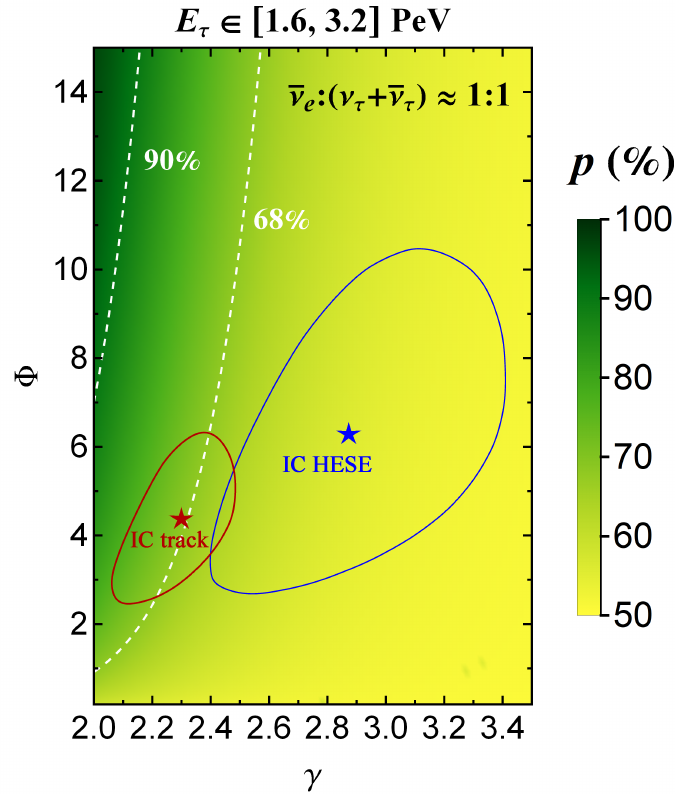}}
		{
			\includegraphics[width=0.28\textwidth]{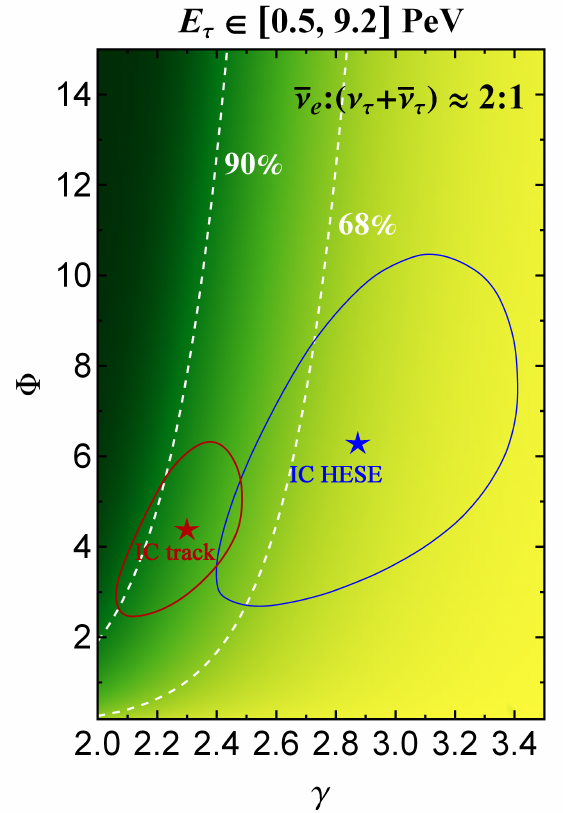}
			\includegraphics[width=0.28\textwidth]{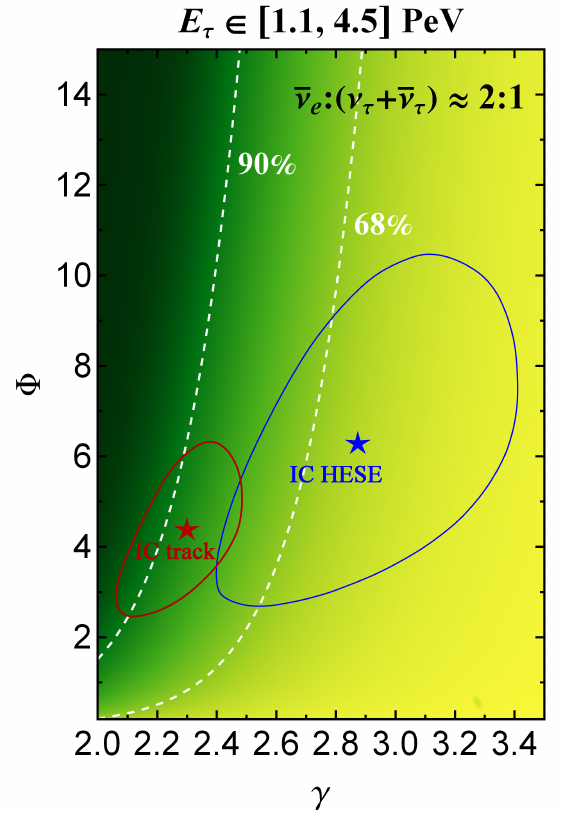}
			\includegraphics[width=0.344\textwidth]{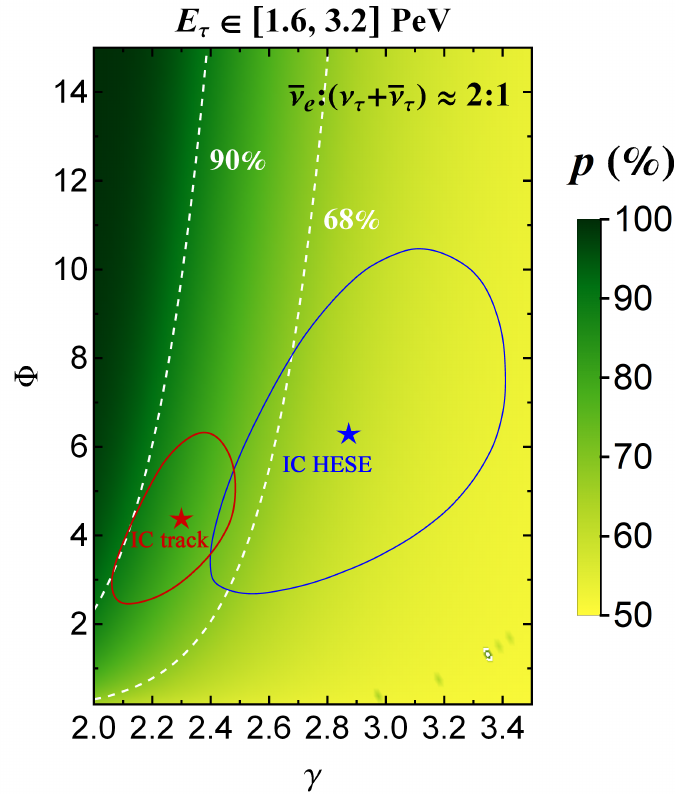}}
	\end{center}
	\vspace{-0.3cm}
	\caption{The $p$-value to exclude the null-signal hypothesis (DIS background only) with different energy intervals in the plane of $\Phi$ and $\gamma$, assuming an exposure of ten years. In the upper and lower panels, the flavor ratio $\nu^{}_{e} + \overline{\nu}^{}_{e} : \nu^{}_{\mu} + \overline{\nu}^{}_{\mu} : \nu^{}_{\tau} + \overline{\nu}^{}_{\tau}$ at Earth has been taken as $1:1:1$ (for  meson decays) and $0.55:0.17:0.28$ (for  neutron decays), respectively. }
	\label{fig:pValue}
\end{figure}

In the upper panels of Fig.~\ref{fig:pValue}, we show the $p$-value in terms of ${\rm \Phi}$ and $\gamma$ for the flavor composition $\nu^{}_{e} + \overline{\nu}^{}_{e} : \nu^{}_{\mu} + \overline{\nu}^{}_{\mu} : \nu^{}_{\tau} + \overline{\nu}^{}_{\tau} = 1:1:1$ consistent with meson-decay neutrino sources, with $\nu^{}_{e}:\overline{\nu}^{}_{e} = 0:1$.
In each panel, the confidence levels of $p = 68\%$ and $90\%$ are indicated by the dashed contours.
If the event number is completely consistent with the background-only hypothesis, one should have $p \approx 50\%$.
The significance by choosing $E^{}_{\tau} \in \left[1.1,\; 4.5\right]~{\rm PeV}$ turns out to be the best among three ROI choices.
However, a $90\%$ CL cannot be achieved for the flux parameters constrained by the IceCube data.
The largest $p$-value we find within the allowed region of neutrino flux is around $82\%$.

One possible way out is the scenario that the incoming tau neutrino component gets suppressed.
An example is the neutron-decay origin of UHE neutrinos, where only $\overline{\nu}^{}_{e}$ is produced at source and after oscillations one roughly has $\nu^{}_{e} + \overline{\nu}^{}_{e} : \nu^{}_{\mu} + \overline{\nu}^{}_{\mu} : \nu^{}_{\tau} + \overline{\nu}^{}_{\tau} = 0.55:0.17:0.28$ with $\nu^{}_{e}:\overline{\nu}^{}_{e} = 0:1$ at Earth~\cite{Song:2020nfh}. 
In this scenario, the $\overline{\nu}^{}_{e}$ flux is nearly twice of the $\nu^{}_{\tau} + \overline{\nu}^{}_{\tau}$ flux.
In the lower panels of Fig.~5, we show the $p$-value of the Glashow resonance discovery for the neutron-decay origin.
One can find that a $90\%$ CL can be reached within the allowed region of neutrino flux.
However, the neutron-decay scenario is not favored as the dominant source of UHE neutrinos because of less energetic neutrinos in the final state.




\section{Concluding remarks} \label{sec:IV}

In the air shower neutrino telescope, the major difficulty to extract the Glashow resonance signal stems from the original tau neutrino events, which contribute as an intrinsic background to the resonance events. 
Even though the Glashow resonance process can induce observable event excess within an energy window,
it is statistically challenging to discriminate the Glashow resonance from the tau neutrino background if UHE neutrinos mainly come from meson decays at  source.
We have identified several factors that limit the discovery potential: (i) the small branching ratio of the decay $W \to \tau\nu^{}_{\tau}$; (ii) the spreading energy distribution of the final-state tau as well as the sizable energy resolution; (iii) the attenuation effect for upward going Earth-skimming neutrinos.
Cosmic tau neutrinos remain the most cost-effective scientific target of the air shower neutrino telescope.
Nevertheless, it is promising to identify the Glashow resonance if UHE neutrinos are of the less likely neutron-decay origin or some new physics can come into play to suppress the tau neutrino component at Earth~\cite{Arguelles:2015dca,Bustamante:2015waa,Arguelles:2022tki}.

Our result above is more or less optimistic. In a more complete analysis, the neutrino flux input itself is uncertain and should be fitted by the data. The flux uncertainty will to some extent dilute the sensitivity to the bump induced by the Glashow resonance.
In the TAMBO-like setup, to disentangle the flux uncertainty one may first fix the tau neutrino spectrum by using the Earth-skimming events ($\alpha \gtrsim 0^{\circ}$), and then use it to generate the background expectation for the horizontal ($\alpha \lesssim 0^{\circ}$) events. 
The measurements by other UHE neutrino telescopes can also be used as the input to further reduce the flux uncertainty.
The TAMBO-like setup may be scaled up by deploying the arrays at multiple sites.
The valleys that have been considered as the host by TAMBO include the Hells Canyon, the Yarlung Tsangpo Grand Canyon, the Cotahuasi Canyon and the Colca Canyon.
Without considering the cost budget, by simply rescaling the length of the toy valley to $1000~{\rm km}$,
a geometric area of $5000~{\rm km^2}$ might be achieved.
With the large statistics, this will make the discrimination of the Glashow resonance possible ($\sim 2\sigma$ CL) even for  meson-decay neutrino sources.

\section*{Acknowledgments}
The author would like to thank Ting Cheng, Sudip Jana and Nele Volmer for helpful comments and discussions.

\bibliographystyle{utcaps_mod}

\bibliography{reference}

\end{document}